\documentclass{iopart}
\usepackage{graphicx}
\usepackage{subfigure}

\begin{document}

\title[Fisher-based thermodynamics for scale-invariant systems]{
Fisher-based thermodynamics for scale-invariant systems: 
Zipf's Law as an equilibrium state of a scale-free ideal gas}

\author{A Hernando$^1$, D Puigdom\`enech$^1$, D Villuendas$^2$ and C Vesperinas$^3$}

\address{$^1$ Departament ECM, Facultat de F\'{\i}sica,
Universitat de Barcelona. Diagonal 647,
08028 Barcelona, Spain}
\address{$^2$ Departament FFN, Facultat de F\'isica, Universitat de Barcelona,
Diagonal 647, 08028 Barcelona, Spain}
\address{$^3$ Sogeti Espa\~na, WTCAP 2, Pla\c ca de la Pau s/n, 08940
Cornell\`a, Spain}

\eads{\mailto{alberto@ecm.ub.es}, \mailto{puigdomenech@ecm.ub.es}, \mailto{diego@ffn.ub.es} and \mailto{cristina.vesperinas@sogeti.com}}

\begin{abstract}
We present a thermodynamic formulation for scale-invariant systems
based on the principle of extreme information. We create an analogy
between these systems and the well-known thermodynamics of gases and
fluids, and study as a compelling case the non-interacting system
---the \emph{scale-free ideal gas}--- presenting some empirical
evidences of electoral results, city population and total cites 
of Physics journals that confirm its existence. The empirical
class of universality known as Zipf's law is derived from first
principles: we show that this special class of power law can be
understood as the density distribution of an equilibrium state of
the scale-free ideal gas, whereas power laws of different exponent
arise from equilibrium and non-equilibrium states. We also predict
the appearance of the log-normal distribution as the equilibrium
density of a harmonically constrained system, and finally derive an
equivalent microscopic description of these systems.
\end{abstract}

\maketitle


\section{Introduction}
\label{p1}

The study of scale-invariant phenomena has unravelled interesting and
somewhat unexpected behaviours in systems belonging to disciplines of
different nature, from physical and biological to technological and
social sciences~\cite{uno}. Indeed, empirical data from percolation
theory and nuclear multifragmentation~\cite{perco} reflect
scale-invariant behaviour, and so do the abundances of genes in
various organisms and tissues~\cite{furu}, the frequency of words in
natural languages~\cite{zip}, scientific collaboration
networks~\cite{cites}, the Internet traffic~\cite{net1}, Linux
packages links~\cite{linux}, as well as electoral
results~\cite{elec1}, urban agglomerations~\cite{ciudad} and firm
sizes all over the world~\cite{firms}.

The common feature in these systems is the lack of a characteristic
size, length or frequency for an observable $k$ at study. This lack
generally leads to a power law distribution $p(k)$, valid in most of
the domain of definition of $k$,
\begin{equation}\label{eq1}
p(k)\sim1/k^{1+\gamma},
\end{equation}
with $\gamma\geq0$. Special attention has been paid to the class of
universality defined by $\gamma=1$, which corresponds to Zipf's law
in the cumulative distribution or the rank-size
distribution~\cite{perco,furu,zip,net1,linux,ciudad,firms,citis}.
Recently, Maillart et al.~\cite{linux} have studied the evolution of the
number of links to open source software projects in Linux packages,
and have found that the link distribution follows Zipf's law as a
consequence of stochastic proportional growth. In its simplest
formulation, the  stochastic proportional growth
 model, or namely the geometric Brownian motion,  assumes the growth of an
element of the system to be proportional to its size $k$, and to be
governed by a stochastic Wiener process. The class $\gamma=1$
emerges from the condition of stationarity, i.e., when the system
reaches a dynamic equilibrium~\cite{citis}.

There is a variety of models arising in different fields that yield
Zipf's law and other power laws on a case-by-case
basis~\cite{ciudad,citis,mod1}. In the context of complex networks,
proportional growth processes known as preferential
attachment~\cite{net1} and competitive cluster growth~\cite{ccg}
have been used to explain many of the properties of natural
networks, from social to biological. The emergence of power laws in
all these models is explained by W. J. Redd and B. D.
Hughes~\cite{exp}, which have shown analytically that models based
on stochastic processes with exponential growth ---as the geometric
Brownian motion, discrete multiplicative process, the
birth-and-death process, or the Galton-Watson branching process---
generate power laws in one of both tails of the statistical
distributions. However, in spite of the success of these models, the
intrinsic complexity involved makes the study at a macroscopic level
difficult since a general formulation of the thermodynamics of
scale-invariant physics is not established yet.

Frieden et al.~\cite{fisher2} have shown that equilibrium and
non-equilibrium thermodynamics can be derived from the principle of
extreme Fisher information. The information measure is done for the
particular case of \emph{translation families}, i.e., distribution
functions whose form does not change under translational
transformations. In this case, Fisher measure becomes
\emph{shift-invariant}~\cite{sym}, what yields most of the canonical
Hamiltonians of theoretical physics~\cite{libro}. Variations of the
information measure lead to a Schr\"odinger equation~\cite{QM} for
the probability amplitude, where the ground state describes
equilibrium physics and the excited states account for
non-equilibrium physics. As for Hamiltonian systems~\cite{H}, it has
been recently shown that the principle of extreme physical
information allows to describe the behaviour of complex systems, as
the allometric or power laws found in biological
sciences~\cite{allometric}.

In this work we present a theoretical framework based on the
principle of extreme physical information that aims to describe
scale-invariant systems at a macroscopic level. We show that the
thermodynamics for such systems can be formulated when the
information measure is taken on distributions that do not change
under \emph{scale} transformations. We show that proportional growth
is intrinsic to this symmetry, and the processes that describe
Zipf's law, as the geometrical Brownian motion, are the equivalent
microscopic description of these systems.

This work is organized as follows. In Sec.~\ref{p2} we present the
Fisher information measure for a scale-invariant system. In Sec.~\ref{p3}
we derive the equilibrium state and non-equilibrium states of the most general
case of non-interacting scale-invariant system: the \emph{scale-free
ideal gas} (SFIG), and present some empirical evidences of its
existence in electoral results and city population. In Sec.~\ref{p4} 
we derive from first principles the special
case of Zipf's law, what we call the \emph{Zipf regime} of the SFIG,
and study empirical data of the total number of cites of Physics
journals to understand the conditions leading to its appearance. 
In Sec.~\ref{p5} we constrain the system
harmonically, finding that the equilibrium density follows a
log-normal distribution. In Sec.~\ref{p6} we derive from the SFIG the
microscopic stochastic equation of motion, showing that the system
can be described by geometrical Brownian walkers. Finally, in
Sec.~\ref{p7} we summarize our results and discuss some aspects of our
work. In the Appendix we derive from the Fisher information the
equations of the well-known translational-invariant ideal gas, which
we use as analogy in the derivation of the SFIG.

\section{The principle of extreme information for a scale-invariant system}
\label{p2}

The Fisher information measure $I$ for a system of $N$ elements,
described by a set of coordinates $\bi{q}$ and physical
parameters $\bi{\theta}$, has the form~\cite{libro}
\begin{equation}\label{fish}
I(F)=\int
\rmd \bi{q}F(\bi{q}|\bi{\theta})\sum_{ij}c_{ij}\frac{\partial}{\partial\theta_i}\ln
F(\bi{q}|\bi{\theta})\frac{\partial}{\partial\theta_j}\ln
F(\bi{q}|\bi{\theta}),
\end{equation}
where $F(\bi{q}|\bi{\theta})$ is the density distribution in
configuration space ($\bi{q}$) conditioned by the physical
parameters ($\bi{\theta}$). The constants $c_{ij}$ account for
dimensionality, and take the form $c_{ij}=c_i\delta_{ij}$ if $q_i$
and $q_j$ are uncorrelated.
Following the principle of extreme information (PEI), the state of
the system extremizes $I$ subject to prior conditions, as the
normalization of $F$ or any constraint on the mean value of an
observable $\langle A_i \rangle$.
The PEI is then written as a variation problem of the form
\begin{equation}
\delta\left\{I(F)-\sum_i\mu_i\langle A_i \rangle\right\}=0,
\end{equation}
where $\mu_i$ are the Lagrange multipliers. In the Appendix, we
derive from the PEI the density distribution in configuration space
and the entropy equation of state for the well-known translational
invariant ideal gas (IG)~\cite{termo}. In analogy with this
derivation, we follow here the same steps to
obtain the SFIG density distributions and entropy equations of state.\\

We consider a one-dimensional system with dynamical coordinates
$\bi{q}=(k,v)$ where $\rmd k/\rmd t=v$. We define $k$ as a discrete
variable, i.e. $k=k_1,k_2,\ldots,k_M$, where $k_i=i\Delta k$ and
$M$, assumed to be large, is the total number of bins of width
$\Delta k$. In order to address the scale-invariance behaviour of $k$
in the Fisher formulation, we change to the new coordinates $u=\ln
k$ and $w=\rmd u/\rmd t$, and assume that $u$ and $w$ are
canonical~\cite{mec} and uncorrelated. This assumption leads to the
proportional growth
\begin{equation}\label{dyn}
\rmd k/\rmd t=v=kw.
\end{equation}
For constant $w$ this equation yields an exponential growth
$k=k_0\rme^{wt}$, which is a uniform linear motion for $u$: $u=wt+u_0$,
with $u_0=\ln k_0$~\footnote{This exponential growth allows to recognize the systems
that we study in this work at the macroscopic level with those
studied in~\cite{exp}.}. It is easy to check that the scale
transformation $k'=k/\theta_k$ leaves invariant the coordinate $w$,
whereas the coordinate $u$ transforms translationally
$u'=u-\Theta_k$, where $\Theta_k=\ln\theta_k$.

If physics does not depend on scale, i.e., the system is
translationally invariant with respect to the coordinates $u$ and
$w$, the distribution of physical elements can be described by the
monoparametric translation families
$F(u,w|\Theta_k,\Theta_w)=f(u',w')$. By analogy with the IG, we
define the SFIG as a system of $N$ non-interacting elements for
which the density distribution can be factorized as
$f(u,w)=g(u)h(w)$. Taking into account that
 $u$ and $w$ are canonical and uncorrelated ($c_{ii}=c_i\neq0$ and $c_{uw}=c_{wu}=0$),
and that the Jacobian for the change of variables is
$\rmd k\rmd v=\rme^{2u}\rmd u\rmd w$, the information measure $I=I_u+I_w$ can be
obtained in the continuous limit as
\begin{equation}
\begin{array}{rl}
I_u=&\displaystyle c_u\int \rmd u~\rme^{2u}g(u)\left|\frac{\partial\ln g(u)}{\partial u}\right|^2\\
I_w=&\displaystyle c_w\int \rmd w~h(w)\left|\frac{\partial\ln h(w)}{\partial w}\right|^2.
\end{array}
\end{equation}

The constraints to the given observables $\langle A_i \rangle$ in
the extremization problem determine the behaviour of the system. In the next sections we
study three different cases: the general case of the
scale-free ideal gas ---the step-by-step analogy of the ideal
gas---, a un-constrained gas or what we call the \emph{Zipf regime}, and the
harmonically constrained gas.

\section{The scale-free ideal gas}
\label{p3}

For the general case, in the extremization of Fisher information we
constrain the normalization of $g(u)$ and $h(w)$ to the total number
of particles $N$ and to $1$, respectively
\begin{equation}\label{w2a}
\int \rmd u~\rme^{2u}g(u)=N,\qquad\int \rmd w~h(w)=1.
\end{equation}
In addition, we penalize infinite values for $w$ with a constraint
on the variance of $h(w)$ to a given measured value
\begin{equation}
\int \rmd w~h(w)(w-\overline{w})^2=\sigma_w^2,\label{w2b}
\end{equation}
where $\overline{w}$ is the average growth. The variation yields
\begin{equation}\label{varg}
\delta\left\{c_u \int \rmd u~\rme^{2u}g\left|\frac{\partial\ln g}{\partial
u}\right|^2+\mu \int \rmd u~\rme^{2u}g\right\}=0
\end{equation}
and 
\begin{equation}\label{varh}
\delta\left\{c_w \int \rmd w~h\left|\frac{\partial\ln
h}{\partial w}\right|^2+\lambda \int \rmd w~h(w-\overline{w})^2 +\nu\int \rmd w~h\right\}=0,
\end{equation}
where $\mu$, $\lambda$ and $\nu$ are Lagrange multipliers.
Introducing $g(u)=\rme^{-2u}\Psi^*(u)\Psi(u)$, and varying
~(\ref{varg}) with respect to $\Psi^*$ leads to the Schr\"odinger
equation
\begin{equation}
\left[-4\frac{\partial^2}{\partial u^2}+4+\mu'\right]\Psi(u)=0,
\end{equation}
where $\mu'=\mu/c_u$. Analogously to the IG, we impose solutions
compatible with a finite normalization of $g$ in the thermodynamic
limit $N,\Omega\rightarrow\infty$ with $N/\Omega=\rho_0$ finite,
where $\Omega=\ln(k_M/k_1)=\ln M$ is the volume in $u$ space and
$\rho_0$ is defined as the \emph{bulk density}. Solutions compatible
with the normalization of~(\ref{w2a}) are given by
$\Psi(u)=A_\alpha \rme^{-\alpha u/2}$, where $A_\alpha$ is the
normalization constant and $\alpha=\sqrt{4+\mu'}$. In this general
case, the density distribution as a function of $k$ takes the form
of a power law: $g_\alpha(\ln
k)=A^2/k^{2+\alpha}$. The equilibrium is defined by the ground state
solution, which correspond the lowest allowed value $\alpha=0$. It
can be show that it is just a uniform density distribution in $u$
space at the bulk density: $g(u)\rme^{2u}\rmd u=N/\Omega \rmd u=\rho_0\rmd u$.

Introducing $h(w)=\Phi^*(w)\Phi(w)$ and varying~(\ref{varh})
with respect to $\Phi^*$ leads to the quantum harmonic oscillator
equation~\cite{QM}
\begin{equation}
\left[-4\frac{\partial^2}{\partial w^2}+\lambda'(w-\overline{w})^2+\nu'\right]\Phi(w)=0,
\end{equation}
where $\lambda'=\lambda/c_w$ and $\nu'=\nu/c_w$. The equilibrium
configuration corresponds to the ground state solution, which is now
a Gaussian distribution. Using~(\ref{w2b}) to identify
$|\lambda'|^{-1/2}=\sigma_w^2$ we get the Boltzmann distribution
\begin{equation}\label{hw}
h(w)=\frac{\exp\left[-(w-\overline{w})^2/2\sigma_w^2\right]}{\sqrt{2\pi}\sigma_w}.
\end{equation}

The density distribution in configuration space
$\widetilde{f}(k,v)\rmd k\rmd v=f(u,w)\rme^{2u}\rmd u\rmd w$ is then
\begin{equation}
\widetilde{f}(k,v)=\frac{N}{\Omega k^2}\frac{\exp\left[-(v/k-\overline{w})^2/2\sigma_w^2\right]}{\sqrt{2\pi}\sigma_w}.\label{fkv}
\end{equation}
If we define $H=\Delta k^2/\Delta\tau$ as the elementary volume in
phase space, where $\Delta\tau$ is the time element, the total
number of microstates is $Z=N!H^N\prod_{i=1}^Nf_1(k_i,v_i)$, where
$f_1=\widetilde{f}/N$ is the monoparticular distribution function
and $N!$ counts all possible permutations for distinguishable
elements. The entropy equation of state $S=-\kappa\ln Z$ reads
\begin{equation}
S=N\kappa\left\{\ln\frac{\Omega}{N}\frac{\sqrt{2\pi}\sigma_w}{H'}+\frac{3}{2}\right\},
\end{equation}
where $\kappa$ is a constant that accounts for dimensionality and
$H'=H/(k_M k_1)=H/(M\Delta k^2)=1/(M\Delta\tau)$. Remarkably, this
expression has the same form as the one-dimensional IG ($D=1$ in
~(\ref{SIG})); instead of the thermodynamical variables
$(N,V,T)$, here we deal with the variables $(N,\Omega,\sigma_w)$,
which make the entropy scale-invariant.

\begin{figure}[t!]
\centering
\subfigure{\includegraphics[width=0.475\linewidth,clip=true]{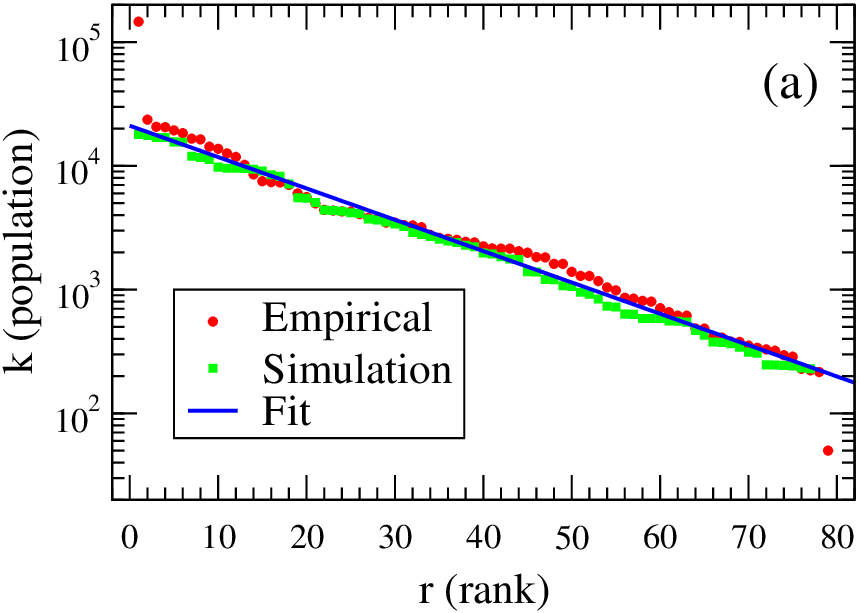}}\hfill
\subfigure{\includegraphics[width=0.475\linewidth,clip=true]{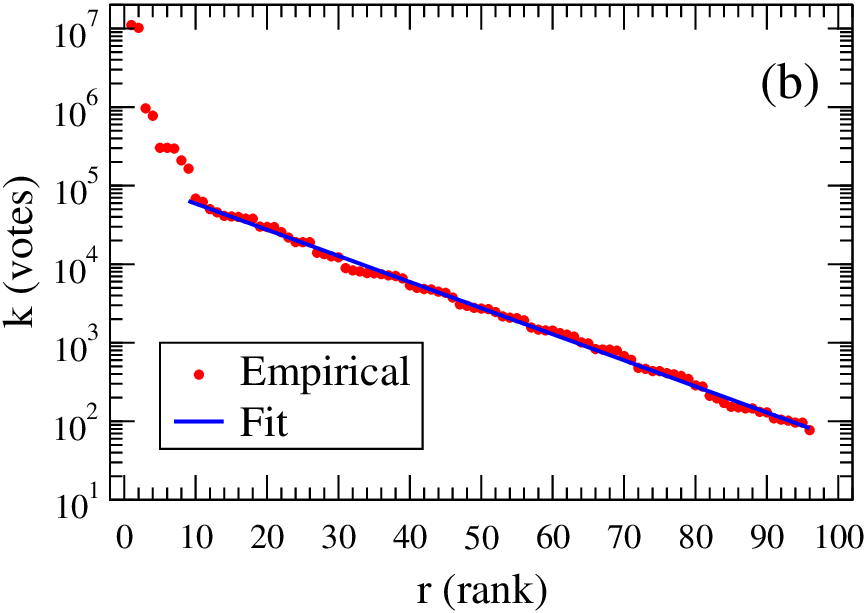}}\\
\subfigure{\includegraphics[width=0.475\linewidth,clip=true]{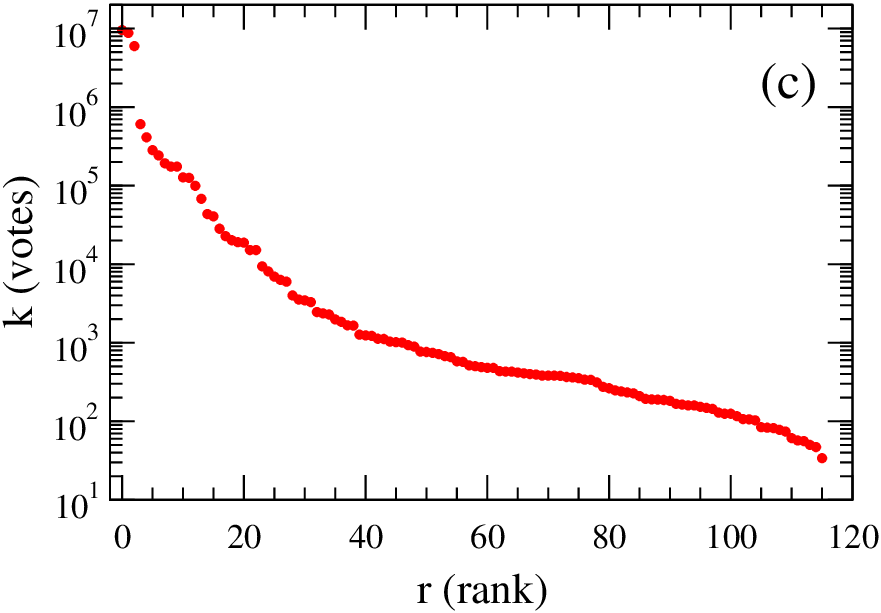}}%
\caption{(colour on-line) \textbf{a}, rank-size distribution of the
cities of the province of Huelva, Spain (2008), sorted from largest
to smallest, compared with the result of a simulation with Brownian
walkers (green squares). \textbf{b}, rank-plot of the 2008 General
Elections results in Spain. \textbf{c}, rank-plot of the 2005
General Elections results in the United Kingdom. (red dots:
empirical data; blue lines: linear fitting).\label{fig1}}
\end{figure}

The total density distribution for $k$ is obtained integrating for
all $v$ the density distribution in configuration space. Integrating
~(\ref{fkv}) we get
\begin{equation}\label{cold}
\widetilde{f}(k)=\int \rmd v\widetilde{f}(k,v)=\frac{N}{\Omega}\frac{1}{k}=\frac{\rho_0}{k},
\end{equation}
which corresponds for large $N$ to an exponential rank-size distribution
\begin{equation}\label{coldr}
k(r)=k_1\exp\left[\Omega-\frac{r-1}{\rho_0}\right],
\end{equation}
where $r$ is the rank. 

This behaviour, which corresponds to the class of universality
$\gamma=0$ in~(\ref{eq1}), has been empirically found by
Costa Filho et al.~\cite{elec1} in the distribution of votes in the
Brazilian electoral results. We have found such a behaviour in the
city-size distribution of small regions and electoral results, like
the province of Huelva (Spain)~\cite{muni}, and the 2008 Spanish
General Elections results~\cite{elecSp}, respectively. We show in
figure~1a and 1b their rank-size distribution in semi-logarithmic
scale, where a straight line corresponds to a distribution
of type~(\ref{coldr}). 
Most of the distribution can be linearly fitted, with a
correlation coefficient of $0.994$ and $0.998$ respectively. 
From these fits we have obtain a bulk density of $\rho_0=0.058$ for the General Elections results,
and in the case of Huelva $\rho_0=17.1$ ($N=77$, $\Omega=4.5$).
Using historical data for the latter~\cite{muni}, we have used the backward differentiation formula to
calculate the relative growth rate of the $i$-th city as
\begin{equation}
w_i = \frac{\ln k_i^{(2008)}-\ln k_i^{(2007)}}{\Delta t}
\end{equation}
where $k_i^{(2007)}$ and $k_i^{(2008)}$ are the number of inhabitants of the
$i$-th city in $2007$ and $2008$ respectively and $\Delta t=1$ year.
We have obtained $\overline{w}=0.012$~years$^{-1}$ and
$\sigma_w=0.032$~years$^{-1}$. However, the regularities are not
always obvious, as shown for the most voted parties in Spain'08 or
the whole distribution of the 2005 General Elections
results in the United Kingdom~\cite{elecUK} (figure~1c). In both cases,
the competition between
parties seems to play an important role, and the assumption of non-interacting 
elements can be unrealistic~\footnote{The effects of interaction are studied in \cite{nota2}, where we go
beyond the non-interacting system using a microscopic description
based on complex networks.}.

\section{The Zipf regime}
\label{p4}

In the previous subsection we considered that $N/\Omega$ remains
finite even in the thermodynamic limit, i.e., the system reaches the
bulk density $\rho_0$. However, if $N/\Omega\rightarrow0$ as
$\Omega\rightarrow\infty$, i.e., the system is exposed to an empty
infinite volume, the normalization can not be achieved and the
constraint has to be removed ($\mu=0$). We call this case the \emph{Zipf regime}, in order to distinguish it from the general.

Considering only the $k$ coordinate in the domain $[k_1,\infty)$, the information measure for the
total density distribution $\widetilde{f}(k)\rmd k=f(u)\rme^{u}\rmd u=p(u)\rmd u$,
reads
\begin{equation}
I_u=c_u\int \rmd u~\rme^{u}f(u)\left|\frac{\partial\ln f(u)}{\partial u}\right|^2,
\end{equation}
and the extremization problem
\begin{equation}\label{vargII}
\delta\left\{c_u \int \rmd u~\rme^{u}f\left|\frac{\partial\ln f}{\partial u}\right|^2\right\}=0.
\end{equation}
Introducing $f(u)=\rme^{-u}\Psi^*(u)\Psi(u)$, and varying with respect
to $\Psi^*$ leads to the Schr\"odinger equation
\begin{equation}
\left[-4\frac{\partial^2}{\partial u^2}+1\right]\Psi(u)=0.
\end{equation}
Taking the boundary conditions $\lim_{u\rightarrow\infty}f(u)=0$ and
$f(u_1)=C$ where $u_1=\ln k_1$ and $C$ is a constant, the solution to the equation is
$\Psi(u)=C' \rme^{-u/2}$, where $C'=\sqrt{C}\rme^{u_1}$. It can be shown
that this is just an exponential decay in $u$ space
$f(u)\rme^{u}\rmd u=C\rme^{-u}\rmd u$. This solution leads to the total density
distribution
\begin{equation}\label{zip}
\widetilde{f}(k)=\frac{C^2}{k^2}
\end{equation}
with $C^2=Nk_1$ for a normalized density. It corresponds in the
continuous limit to a rank-size distribution of the type
\begin{equation}\label{zipr}
k(r)=\frac{N k_1}{r},
\end{equation}
which is the Zipf's law (universal class $\gamma=1$) of
~\cite{perco,furu,zip,net1,linux,ciudad,firms,citis}.
This result is remarkable: for the first time \emph{Zipf's law is derived from first
principles}.

In figure~2 we show the known behaviour~\cite{citis} of the rank size
distribution of the top 100 largest cities of the United
States~\cite{usa}, which shows an slope near $-1$ ($\gamma=1$) in
the logarithmic representation of the rank-plot.


\begin{figure}[t!]
\center
\includegraphics[width=0.5\linewidth,clip=true]{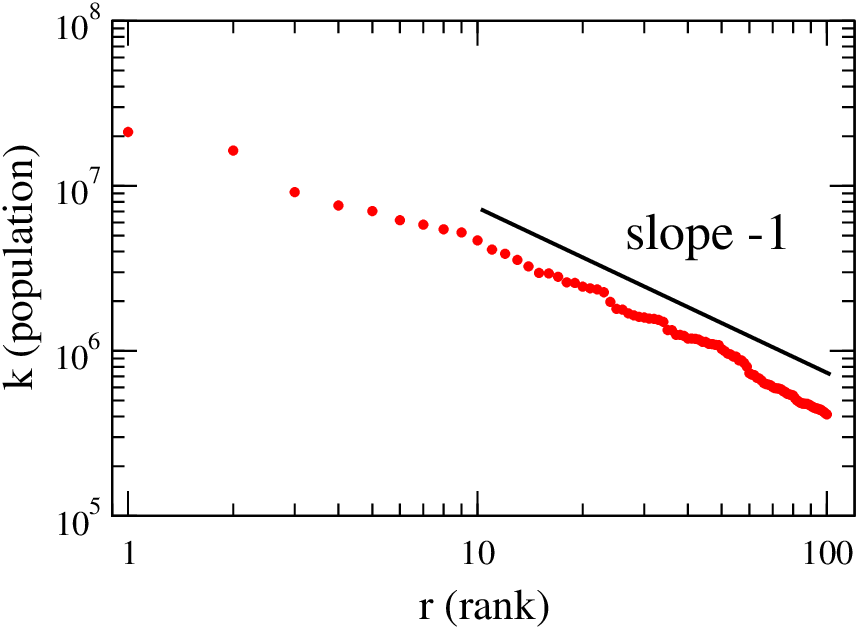}
\caption{(colour on-line) Rank-plot of the 100 largest
cities of the United States.\label{fig2}}
\end{figure}

The appearance of the bulk and the Zipf regime in a SFIG can be
understood studying empirical data. We have studied the system
formed by all Physics journals~\cite{isi} ($N=310$) using their total number of cites as
coordinate $k$. If a journal receives more cites due to its popularity, it
becomes even more popular and therefore it will receive more cites. Under such conditions
proportional growth and
scale invariance are expected. Since we consider all fields of
Physics, correlation effects are much lower than only consider
journals of an specific field, so the non-interacting approximation
seems realistic in this case. In figure~3 we show the rank-plot of the
number of cites of Physic journals, where we have found a slope
near $-1$ for the most-cited journals in the logarithmic
representation (figure~3a) and an slope near $+1$ for the less-cited
journals (figure~3b). For the central part of the distribution bulk density reaches a
value of $\rho_0\sim57$ (figure~3c).

\begin{figure}[t!]
\centering
\subfigure{\includegraphics[width=0.475\linewidth,clip=true]{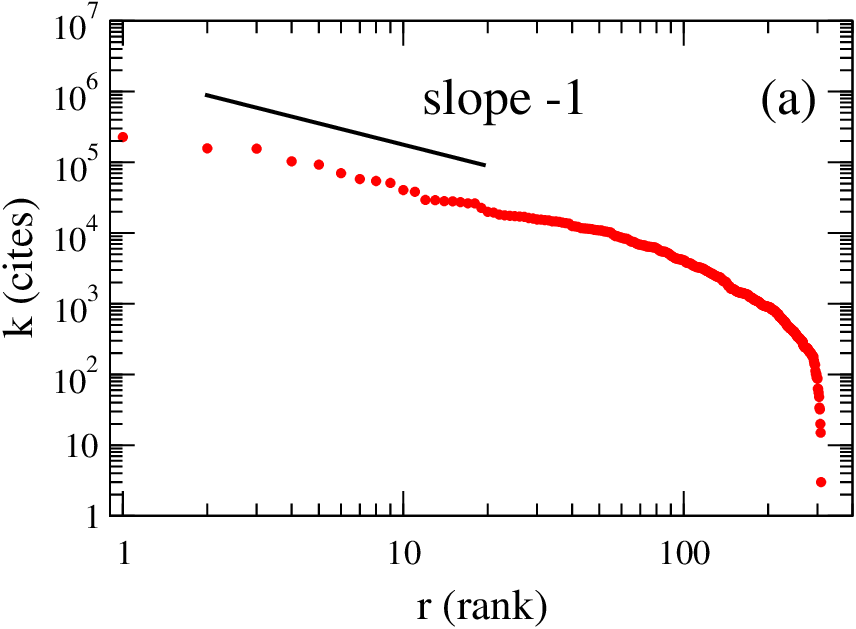}}\hfill
\subfigure{\includegraphics[width=0.475\linewidth,clip=true]{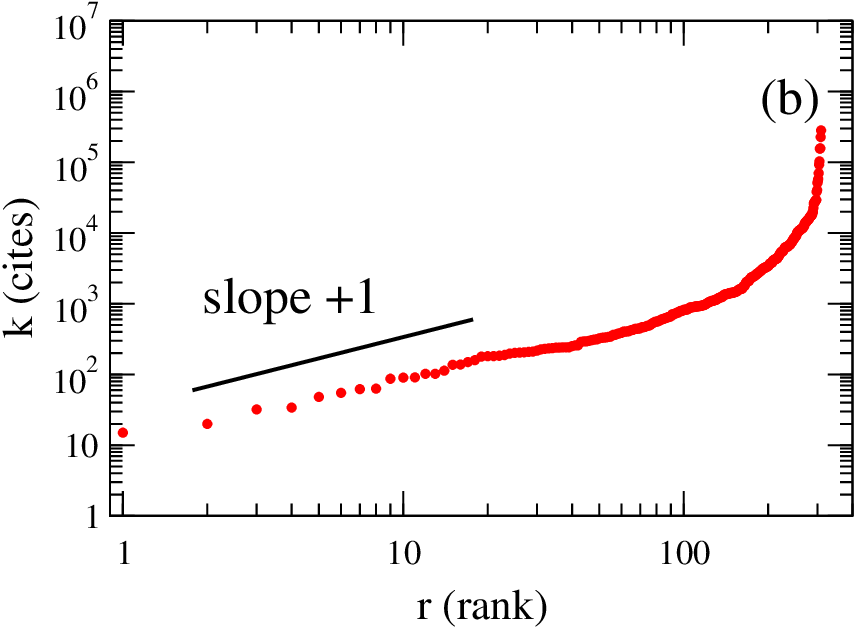}}\\
\subfigure{\includegraphics[width=0.475\linewidth,clip=true]{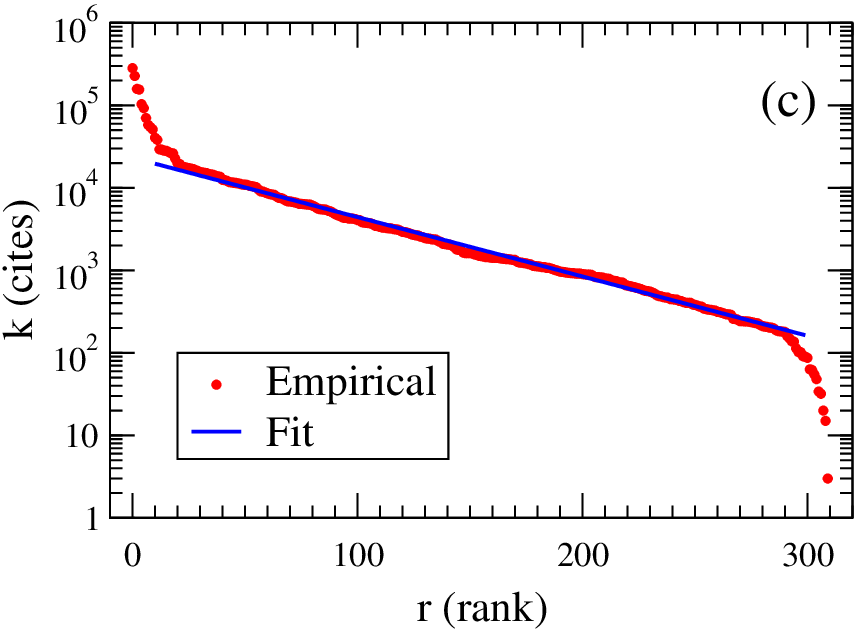}}%
\caption{(colour on-line) \textbf{a}, rank-plot of the total number
of cites of physics journal, from most-cited to less-cited, in logarithm scale.
\textbf{b}, sorted from less-cited to most cited  \textbf{c}, same as a,
in semi-logarithm scale.(red dots: empirical data;
blue line: linear fitting).\label{fig3}}
\end{figure}

This distribution shows an extraordinary symmetric behaviour under
the change $k\rightarrow1/k$ ($u\rightarrow-u$). 
We show in figure~4 the raw empirical data
compared with the distribution obtained from the transformation
$k'=c/k$ ($u'=-u+\ln c$), where $c=3.3\times10^6$. The symmetry of this
system is an important clue to understand both regimes, and
represents a perfect example of the conditions needed to observe
bulk and Zipf regimes in a non-interacting scale-invariant system.
The main part of the density distribution reaches the bulk density
obeying~(\ref{cold}), whereas Zipf's law emerge at the edges,
obeying~(\ref{zip}): following the analogy with the physics of
gases and fluids, we can think of the system as a drop, where the
Zipf regime is the sign of a \emph{surface} since it reproduces how
the density exponentially falls from the bulk density to zero in $u$
space when the system is exposed to an infinite empty volume. This
effect is clearly visible in figure~5, where the empirical density 
distribution $p(u)\rmd u$ in $u$ space is compared with the fitted density
\begin{equation}\label{fit}
p(u)=\left\{
\begin{array}{ll}
\rho_a\rme^{u}&\mathrm{if~} u<u_a\\
\rho_0&\mathrm{if~} u_a<u<u_b\\
\rho_b\rme^{-u}&\mathrm{if~} u_b>u
\end{array}
\right.
\end{equation}
where $\rho_a=0.1$, $\rho_0=57$, $\rho_b=4.3\times10^5$, $u_a=5.2$ and $u_b=10$.
These findings lead us to conclude that the system of Physics journals
sorted by total number of cites is a perfect example of the scale-free
ideal gas at equilibrium.

\begin{figure}[t!]
\center
\includegraphics[width=0.5\linewidth,clip=true]{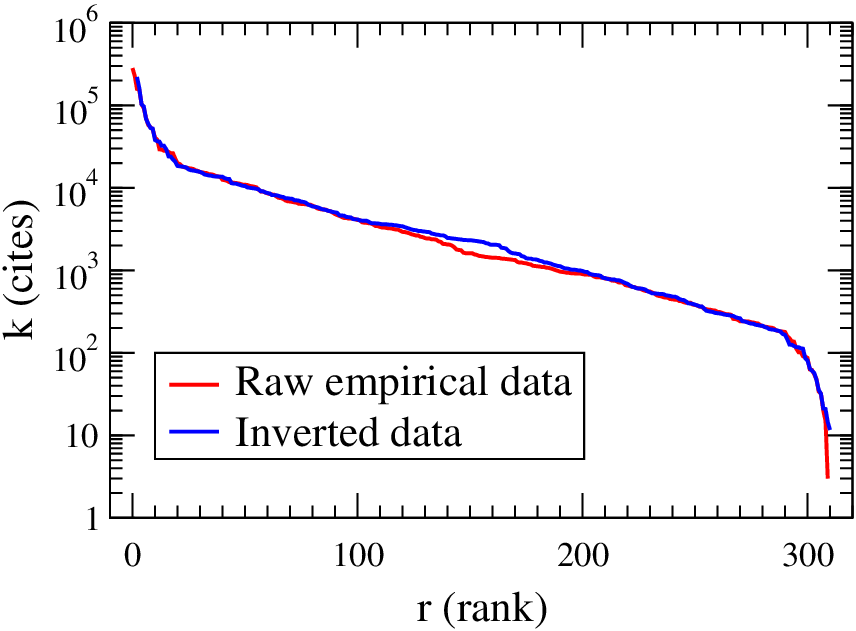}
\caption{(colour online) Rank-plot of the total number of cites of
Physics journal, from most-cited to less-cited, compared with the
distribution obtained from the inverse transformation $k'=3.3\times10^6/k$
where $k$ is the number of cites.\label{fig4}}
\end{figure}

\begin{figure}[t!]
\center
\includegraphics[width=0.5\linewidth,clip=true]{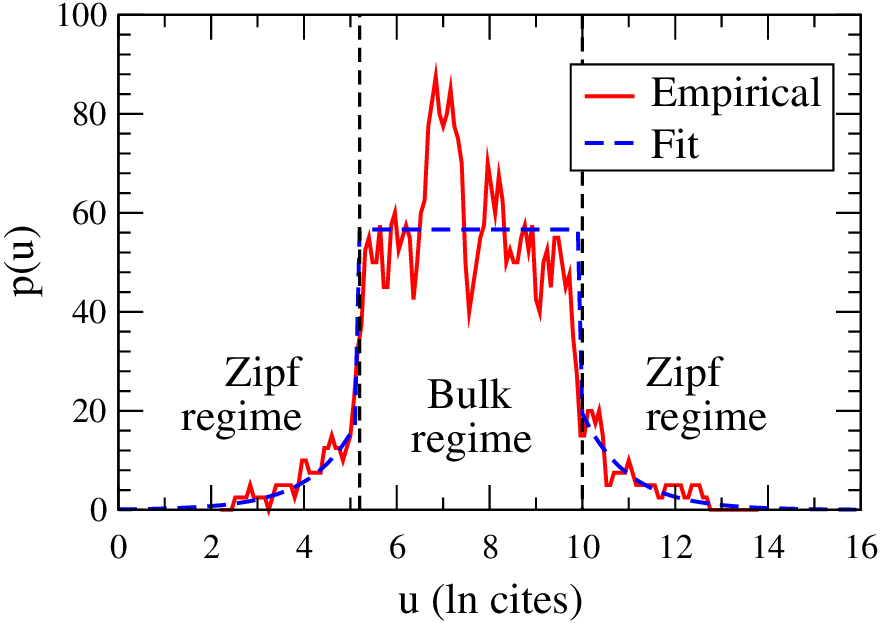}
\caption{(colour online) Empirical density distribution in $u$ space
of the total number of cites of Physics journals, compared with
~(\ref{fit}). The bulk regime and the Zipf regime at the
edges is clearly visible.
\label{fig5}}
\end{figure}

\section{The harmonically constrained system}
\label{p5}

We now consider a system with a constraint in a given observable
$\langle A \rangle$ which locally depends on $k$, $A=A(k)$.
The second order Taylor expansion with respect to $u=\ln k$
near a minimum is written as $\widetilde{A}(u)=A(\rme^u)\simeq
A_0+A_2/2(u-u_m)^2$, where $A_0$, $A_2$ and $u_m$ are constants.
Introducing this constraint and the normalization condition to the number of elements $N$
of the total density distribution, the extremization problem reads
\begin{equation}\label{vargIII}
\begin{array}{rl}
\delta&\displaystyle\left\{c_u \int \rmd u~\rme^{u}f\left|\frac{\partial\ln f}{\partial u}\right|^2+\mu \int \rmd u~\rme^{u}f\right.\\
&\displaystyle\left.+\lambda \int \rmd u~\rme^{u}f\left[A_0+\frac{1}{2}A_2(u-u_m)^2\right]\right\}=0
\end{array}
\end{equation}
Introducing $f(u)=\rme^{-u}\Psi^*(u)\Psi(u)$, and varying with respect
to $\Psi^*$ leads to the quantum harmonic oscillator equation
\begin{equation}
\left[-4\frac{\partial^2}{\partial u^2}+\lambda'(u-u_0)^2+\nu'\right]\Psi(u)=0,
\end{equation}
where now we have defined $\lambda'=(\lambda A_2)/2c_w$ and
$\nu'=(\nu+1+\lambda A_0)/c_u$. The ground state solution is a
gaussian distribution, which now yields a total density
distribution of the form of a log-normal distribution
\begin{equation}
\widetilde{f}(k)=\frac{N}{k\sqrt{2\pi}\sigma_u}\exp\left(\frac{-(\ln k-u_m)^2}{2\sigma_u^2}\right),
\end{equation}
with $|\lambda'|^{-1/2}=\sigma_u^2=\langle A \rangle-A_0$. Note that
if $A_0=0$, the constraint can be also understood as a constraint in the variance of
$u$.

The log-normal distribution has been widely observed in a large
number of scale-invariant systems~\cite{logn}. In~\cite{elec2}
S. Fortunato and C. Castellano found this behaviour in the electoral
results of different countries and for different years. We can think of this
constraint as the effect of polices or social factors: low
popularity candidates are penalized since the party does not present
them for the elections, and high popularity candidates are penalized
by the competition in campaign. Both effects can be approximated to
second order as a harmonic potential, however anharmonic effects are
expected in a high order study.

Defining $H''=1/\Delta\tau$ and $k_m=\ln u_m$ being $k_m=m\Delta k$,
the entropy equation of state reads in this case
\begin{equation}
S=N\kappa\left\{\ln\frac{2\pi\sigma_u\sigma_w m^2}{N H''}+2\right\},
\end{equation}
which maintains scale invariance.

\section{The microscopic description}
\label{p6}

The dynamics of the system can be microscopically described as a
stochastic process using~(\ref{dyn}) and the density
distribution~(\ref{hw}). Treating $w$ as a random variable, the
stochastic equation of motion is written as a geometrical Brownian
motion
\begin{equation}\label{eqd}
\rmd k = k\overline{w}\rmd t + k\sigma_w\rmd W,
\end{equation}
where $\rmd W$ is a Wiener process. In the $u$ space, this equation
reads
\begin{equation}\label{eqd2}
\rmd u = \overline{w}\rmd t + \sigma_w\rmd W,
\end{equation}
which describes the well-known Brownian motion.~(\ref{eqd})
exactly describes the dynamical condition found empirically in
~\cite{linux} and also the stochastic proportional
growth model used in~\cite{citis} to obtain Zipf's law. 
We can think of this sort of simulations as the
equivalent of molecular dynamics simulations for gases and
liquids~\cite{DM}.

Effectively,~(\ref{eqd2}) implies that a uniform density in $u$
space of $N$ Brownian walkers moving in a fixed volume $\Omega$ ---a
model used in the literature to describe the IG~\cite{DM}---
describes the SFIG when we represent the system with the coordinates
$(k,v)$. In figure~1a we show the rank-plot for a system of $N=78$
geometrical Brownian walkers with $\sigma_w=0.029$ in a volume of
$\Omega=4.5$ and $k_1=200$ in reduced units, which nearly describes
the distribution of the population of the province of Huelva.

\section{Summary and Discussion}
\label{p7}

We have shown that a thermodynamic description of scale-invariant systems can
be formulated from the principle of extreme information, finding an
analogy with the thermodynamics of gases and fluids. We have derived
the density distribution in configuration space and the entropy
equation of state of the scale-free ideal gas in the thermodynamic
limit, and have found empirical evidences of its existence in city
population, electoral results and cites to Physics journals. In
this context, Zipf's law emerges naturally as the equilibrium
density of the non-interacting system when the volume grows to
infinity, what we call the Zipf regime. Using empirical
data we have seen that this regime can be understood as the density fall of a
surface between the bulk and an empty volume. We have also studied
the effect of a harmonic constraint, finding that in this case the density of the
system follows a log-normal distribution, which has
been empirically observed in electoral results and in many other
scale-invariant systems~\cite{logn}. Finally we have shown with a simulation of
city population that a geometrical Brownian motion can describe the
system at a microscopic level.

It is well known that in real gases the most interesting situations
emerge when interactions between particles become relevant,
originating deviations from the equation of state of the IG, and
making room for the appearance of, e.g., phase transitions~\cite{termo}.
Analogously, one should also expect this rich phenomenology to show
up in scale-invariant real systems, which may explain deviations
from Zipf's law in empirical distributions. A study beyond the ideal
gas is in progress, and further results will be reported~\cite{nota2}.

\ack 
We would like to thank M. Barranco, R. Frieden,  A. Plastino, and B.
H. Soffer for useful discussions. This work has been partially
performed under grant FIS2008-00421/FIS from DGI, Spain (FEDER).

\appendix

\section{The translational invariant ideal gas}
\label{aa}

In this appendix we derive from the principle of extreme information
the density distribution in configuration space and the entropy equation of state
of the translational invariant ideal gas (IG)~\cite{termo}.
The IG model describes non-interacting classical particles of mass
$m$ with coordinates $\bi{q}=(\bi{r},\bi{p})$, where
$m\rmd \bi{r}/\rmd t=\bi{p}$. We assume that these coordinates are
canonical~\cite{mec} and uncorrelated. This assumption is
introduced in the information measure~(\ref{fish}) as
$c_{ij}=c_i\delta_{ij}$, where $c_{i}=c_r$ for space coordinates,
$c_{i}=c_p$ for momentum coordinates, and $\delta_{ij}$ is the
Kronecker delta. The density distribution can be factorized as
$f(\bi{r},\bi{p})=\rho(\bi{r})\eta(\bi{p})$, and the
information measure $I=I_r+I_p$ reads, if $D$ is the dimension of the space
\begin{equation}
\begin{array}{rl}
I_r=&\displaystyle c_r\int
\rmd^D\bi{r}~\rho(\bi{r})\left|\bi{\nabla}_r \ln \rho(\bi{r})\right|^2\\
I_p=&\displaystyle c_p\int \rmd^D\bi{p}~\eta(\bi{p})\left|\bi{\nabla}_p\ln
\eta(\bi{p})\right|^2.
\end{array}
\end{equation}

In the extremization of Fisher information we constrain the
normalization of $\rho(\bi{r})$ and $\eta(\bi{p})$ to the
total number of particles $N$ and to $1$, respectively
\begin{equation}\label{normg}
\int \rmd^D\bi{r}~\rho(\bi{r})=N,\qquad\int \rmd^D\bi{p}~\eta(\bi{p})=1.
\end{equation}
In addition, we penalize infinite values for the particle momentum
with a constraint on the variance of $\eta(\bi{p})$ to a given
measured value
\begin{equation}\label{sp2}
\int \rmd^D\bi{p}~\eta(\bi{p})(\bi{p}-\overline{\bi{p}})^2=D\sigma_p^2,
\end{equation}
where $\overline{\bi{p}}$ is the mean value of $\bi{p}$. For
each degree of freedom it is known from the Virial theorem that the
variance is related to the temperature $T$ as $\sigma_p^2=mk_BT$,
being $k_B$ the Boltzmann factor. The variation yields
\begin{equation}\label{exg}
\displaystyle\delta\left\{c_r\int \rmd^D\bi{r}~\rho\left|\bi{\nabla}_r \ln
\rho\right|^2+\mu\int \rmd^D\bi{r}~\rho\right\}=0
\end{equation}
and
\begin{equation}\label{exh}
\delta\left\{c_p\int \rmd^D\bi{p}~\eta\left|\bi{\nabla}_p\ln \eta\right|^2+\lambda\int \rmd^D\bi{p}~\eta(\bi{p}-\overline{\bi{p}})^2 +\nu\int
\rmd^D\bi{p}~\eta\right\}=0,
\end{equation}
where $\mu$, $\lambda$ and $\nu$ are Lagrange multipliers.

Introducing
$\rho(\bi{r})=\Psi^*(\bi{r})\Psi(\bi{r})$ and varying~(\ref{exg}) with
respect to $\Psi^*$ leads to the Schr\"odinger equation~\cite{QM}
\begin{equation}
\left[-4\nabla_r^2+\mu'\right]\Psi(\bi{r})=0,
\end{equation}
where $\mu'=\mu/c_r$. To fix the boundary conditions, we first
assume that the $N$ particles are confined in a box of volume $V$,
and next we take the thermodynamic limit (TL)
$N,V\rightarrow\infty$ with $N/V$ finite. The equilibrium state
compatible with this limit corresponds to the ground state solution, which
is the uniform density $\rho(\bi{r})=N/V$.

Introducing $\eta(\bi{p})=\Phi^*(\bi{p})\Phi(\bi{p})$
and varying~(\ref{exh}) with respect to $\Phi^*$ leads to the
quantum harmonic oscillator equation~\cite{QM}
\begin{equation}
\left[-4\nabla_p^2+\lambda'(\bi{p}-\overline{\bi{p}})^2+\nu'\right]\Phi(\bi{p})=0,
\end{equation}
where $\lambda'=\lambda/c_p$ and $\nu'=\nu/c_p$. The equilibrium
configuration corresponds to the ground state solution, which is now a
gaussian distribution. Using~(\ref{sp2}) to identify
$|\lambda'|^{-1/2}=\sigma_p^2$ we get the Boltzmann distribution,
which leads to a density distribution in configuration space of the
form
\begin{equation}
f(\bi{r},\bi{p})=\frac{N}{V}\frac{\exp\left[-(\bi{p}-\overline{\bi{p}})^2/2\sigma_p^2\right]}{(2\pi\sigma_p^2)^{D/2}}.
\end{equation}
If $H$ is the elementary volume in phase space, the total number of
microstates is
$Z=N!H^{DN}\prod_{i=1}^Nf_1(\bi{r}_i,\bi{p}_i)$, where
$f_1=f/N$ is the monoparticular distribution and $N!$ counts all
possible permutations for distinguishable particles. The entropy
$S=-k_B\ln Z$ is written as
\begin{equation}\label{SIG}
S=Nk_B\left\{\ln\frac{V}{N}\left(\frac{2\pi
\sigma_p^2}{H^2}\right)^{D/2}+\frac{2+D}{2}\right\},
\end{equation}
where we have used the Stirling approximation for $N!$. This expression
is in exact accordance with the known value of the entropy for the
IG~\cite{termo}, which shows the predictive power of the Fisher
formulation.\\

\end{document}